# Near-optimal intense and powerful terahertz source by optical rectification in lithium niobate crystal

**L. GUIRAMAND,[1] J. E. NKECK,[1] X. ROPAGNOL,[1,2] T. OZAKI,[2] AND F. BLANCHARD[1,*]**

[1] *Département de génie électrique, École de technologie supérieure, Montréal, Québec, Canada*
[2] *Institut National de la Recherche Scientifique – Énergie Matériaux Télécommunications, Varennes, Québec, Canada*
*Corresponding author: francois.blanchard@etsmtl.ca*

**Using an affordable ytterbium laser with sub-mJ of energy combined with a novel pulse compression technique, we demonstrate an extremely competitive state-of-the-art terahertz (THz) source with 53 mW of average power and 310 kV/cm at focus from the tilted-pulse front pumping scheme in lithium niobate at room temperature. Key points of this demonstration include the use of a pump pulse duration of 280 fs in combination with an echelon mirror. Our results present unmatched combined characteristics and are highly competitive with the existing THz sources pumped at the mJ range. This demonstration is a step towards the democratization of access to intense and powerful THz pulses.**

## 1. INTRODUCTION

Intense electromagnetic radiation in the terahertz (THz) frequency range has been used for various groundbreaking scientific demonstrations [1-8]. The common denominator of these discoveries resulting from light-matter interactions is the development of intense THz pulse sources [9]. Among the various methods for generating THz pulses, the tilted-pulse front pumping (TPFP) scheme in lithium niobate (LN) is a recognized method to produce single-cycle THz pulses with high intensity [10,11] and still the subject of intense research activity with new developments reported every year [12-23]. Due to its very large second-order nonlinear coefficient [11] and its high damage energy threshold [24], LN is the material of choice for laser pumping pulse energies beyond tens of mJ [15, 16, 18, 21-23]. Thanks to these unique laser facilities, some of the highest THz pulse energies reported to date have reached 0.125 mJ [16], 0.2 mJ [21], 0.4 mJ [18], and 1.4 mJ [22] for an optical pump energy of 45 mJ, 70 mJ, 58 mJ and 214 mJ, respectively. Importantly, these THz sources are very intense but not necessarily of high average power due to the low repetition rate of the pump laser.

In parallel with these developments of high-intensity THz pulse sources, a new trend is beginning to emerge in the literature: sources with high average output power capability [25-27]. Recently, results have shown THz pulse sources with average powers up to 66 mW [25] and 144 mW [26]. These demonstrations coincide with the rise of ytterbium (Yb) lasers as a promising replacement for the Ti:Sapphire lasers for THz generation. Compared to the Ti:Sapphire laser, the Yb laser maintains a higher average power in its regenerative amplification section [28], thus providing high repetition-rate output laser pulses. The main disadvantage of this type of laser is the longer pulse width than that obtained with Ti:Sapphire lasers, which can however be advantageous for THz generation using the LN crystal [17]. Nevertheless, driving LN material with very high average power lasers introduces new challenges, such as heat dissipation in LN crystal [26]. So far, for these high repetition rate sources, only low energy conversion efficiencies below 0.1% and with moderate peak THz fields of <150 kV/cm have been reported [25-27]. Ideally, both characteristics must be met simultaneously: high intensity to access the nonlinear light-matter interactions or high source brightness and high average power to allow sensitive detection of the observed phenomena. However, to the best of our knowledge, this combined functionality is so far an unmet goal.

In this work, we report an intense THz pulse source with 310 kV/cm at focus with an average power of 53 mW. Instead of using a grating to perform the TPFP method [10], we used an echelon mirror, which is particularly easy to implement compared to the former. Theoretical predictions have emphasized the importance of this type of device over gratings for achieving longer interaction length [29]. So far, this method has only been demonstrated using an ultra-short (<100 fs) pump pulse [20] and at low pump power using a digital micromirror device [30], but without achieving better performance than the TPFP scheme using a grating. The high performance reported in this work with a near-optimal room temperature conversion efficiency of 1% capitalizes on a long Fourier-limited pump pulse duration, as experimentally anticipated [17] and the long interaction length of the optical rectification process in LN crystal, as theoretically predicted when the tilted-pulse-front is obtained from an echelon mirror [29]. The demonstration of these results was possible through the use of a novel technique of probe pulse compression using only 1 μJ from the fundamental beam at 1.0 μm wavelength. Namely, an ultrashort 512 nm probe pulse with a duration of 75 fs is obtained by first broadening the spectrum by self-phase modulation in a pair of cadmium sulfide (CdS) crystals and spatially filtering, using the second harmonic in a beta barium borate (BBO) crystal, the different modes generated by the broadening process in the CdS [J. E. Nkeck et al., manuscript in preparation]. Our results represent unmatched combined characteristics with pump beam energy as low as 208 μJ.

## 2. EXPERIMENTAL SETUP

### A. Generation section

Figure 1 shows the experimental setup used to demonstrate this high intensity and powerful THz generation using an Yb solid-state amplified laser (model Pharos: PH1-10W from Light Conversion). Its central wavelength is 1.024 μm with a bandwidth of 6.1 nm for a pulse duration of 280 fs. The maximum energy is 400 μJ with a repetition rate of 25 kHz at an average laser power of 10 W. The repetition rate can be adjusted from 25 kHz to 200 kHz while maintaining a maximum optical output power of 10 W. In our configuration, a beam splitter ensures 95% reflection for the pump beam and 5% transmission for the probe beam. The pump beam is sent directly perpendicular to the surface of a Stavax, Ni-P stepped mirror (from Sodick F.T. Co.) with an aperture of 20 mm x 20 mm. The step size of the echelon mirror is 150 μm wide and 75 μm high. The image of the echelon mirror on the LN crystal is obtained with an achromatic cylindrical lens of 100 mm focal length in the x-direction. The generator crystal is made of stoichiometric LN with ~1% MgO doping concentration and consists of a prism shape cut at an angle of 63°. In order to satisfy the phase-matching condition between the pump pulse and the emitted THz pulse, the tilt image from the echelon mirror is demagnified by a factor of ~4.3. This demagnification factor was calculated using equation (1) in ref. [20].

The emitted THz radiation is collected 60 mm away from the LN crystal by a 1-inch focal length, 1-inch diameter gold-coated off-axis parabolic mirror (OAPM). It is then collimated and refocused onto the electro-optical detector crystal using a pair of gold-coated OAPM with 2 inches in diameter and 6 and 2 inches in focal length, respectively. This configuration increases the THz beam diameter by a factor of 6 and thus increases the focusing capability at the detection position.

### B. Detection section

The detection of THz pulses has been done by electro-optic (EO) sampling [31] in a 19 μm thick x-cut LN crystal deposited on a 1 mm thick z-cut LN substrate (fabricated by NanoLN). This detector has some advantages: (i) being very thin, the active layer allows detection over a wide range of frequencies without the problem of phase matching between the optical probe and the THz wave [32]; (ii) the substrate delays the first echo of the THz pulses by more than 40 ps, thus allowing to increase the spectral resolution and; (iii) when working with intense THz pulses, a thin crystal reduces the overall modulation of the EO effect, which allows to operate in a linear response range without the use of an attenuation filter such as high-resistivity silicon wafers.

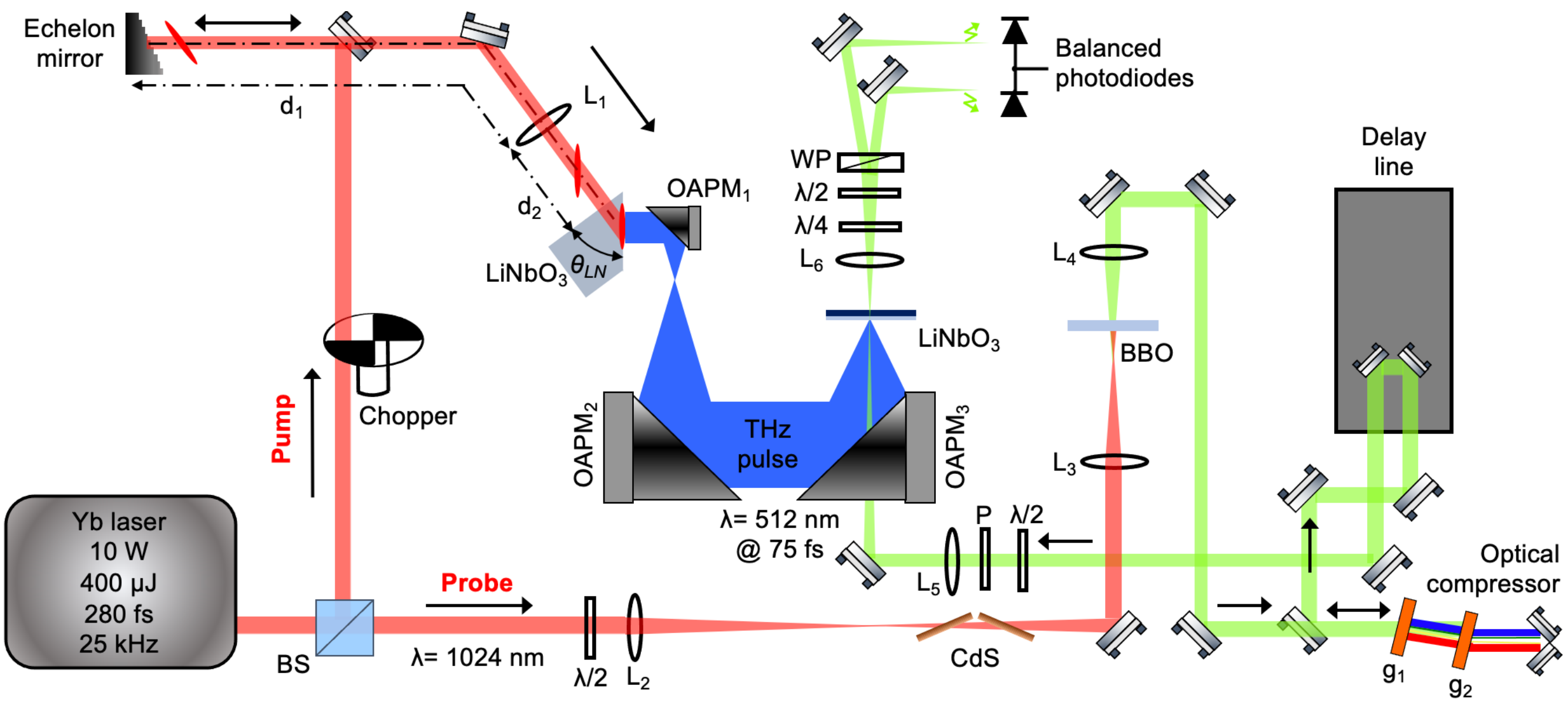

**Fig. 1.** Experimental setup for the generation and the detection of THz pulses with the LN and their detection by EO sampling. BS = beam splitter; $d_1$=550mm; $d_2$=125mm; $\theta_{LN}$ = LN cut angle of 63° ; $L_1$= 100 mm focal lens; $L_2$= 300 mm focal lens; $L_3$= 50 mm focal lens; $L_4$= 75 mm focal lens; $L_5$= 100 mm focal lens; $L_6$= 150 mm focal length lens; $OAPM_1$= off-axis parabolic mirror of 1 inches reflected focal length; $OAPM_2$= off-axis parabolic mirror of 6 inches reflected focal length; $OAPM_3$= off-axis parabolic mirror of 2 inches reflected focal length; $g_1$ and $g_2$= transmissive diffracting gratings with 300 grooves/mm; λ/2 = half-wave plate; λ/4 = quarter wave plate; WP = Wollaston prism; P = polarizer.

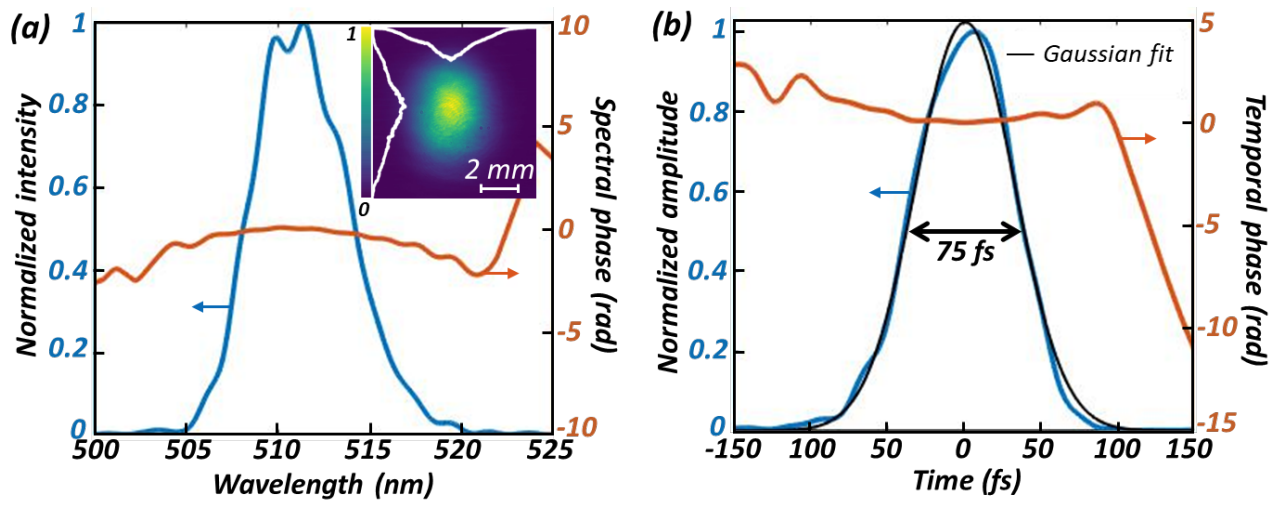

**Fig. 2.** Measured characteristics of the probe pulse after temporal compression. (a) Spectral amplitude and spectral phase distribution with the image of the probe spot in the inset. (b) Temporal intensity and temporal phase distribution.

To optimally detect the electric field of a THz wave by EO sampling, it is essential to work with probe pulses of about 100 fs or less [33]. Therefore, the Pharos laser pulse duration of 280 fs is too long to effectively detect our OR-generated THz pulses and must be compressed. To do so, we used an original laser pulse compression technique, as detailed in reference [J. E. Nkeck et al., manuscript in preparation]. This method requires only a micro-joule or less of probe energy at the fundamental wavelength of the laser, i.e., at 1.024 µm. This laser compression method consists of three distinct steps: (i) pulse spectrum broadening by phase modulation, (ii) spatial and spectral filtering in a thin nonlinear crystal by second harmonic generation, and (iii) laser pulse compression using a pair of transmission gratings.

For spectral broadening by self-phase modulation (SPM), the P-polarized probe pulse is focused into a pair of 1-mm-thick CdS crystals placed at Brewster angle with opposite orientation (see the schematic in Fig. 1). In our case, the spectrum of the probe pulse at FWHM, after the CdS crystals, is broadened from 6.1 nm to 22 nm. Then, this probe beam is focused in to a 100 µm thick type I BBO crystal to generate the second harmonic at a wavelength of 512 nm. The BBO crystal has two main functions: to spatially filter the probe beam by adjusting the position of the BBO crystal with respect to the focus of the optical beam and to select a wide and uniform spectral range. This probe pulse is finally compressed using a pair of standard transmissive gratings (GT25-03 from Thorlabs), which is also used as a geometrical filter of the remaining fundamental beam. It is worth mentioning that the use of an optical probe at the 512 nm wavelength significantly improves the dynamic range of the detection since the detected electric field is inversely proportional to the wavelength [31].

Figure 2 shows (a) the spectral amplitude and spectral phase of the probe pulse measured using FROG and in (b) the temporal intensity and temporal phase of the compressed pulse after transmission through the gratings. The spectral bandwidth at FWHM of the probe beam is 6 nm and the duration at FWHM is 75 fs. We also observe that the spectral and temporal phase is almost flat and the temporal intensity profile of the pulse is Gaussian (i.e., without pedestal), demonstrating the good performance of our probe pulse compression chain.

### C. Pump beam characteristics

A key point of the TPFP technique for THz pulse generation is to correctly image, on the LN crystal, the optical element that tilts the pulse wavefront. Ideally, the greater the interaction in the nonlinear material, the more efficient the OR process will be. This statement is only true if the pulse retains its spatial and temporal properties. Unfortunately, when using a diffraction grating, it has long been known that degradation of the pulse duration away from the image plane occurs due to angular dispersion [34] and that the image of the tilted pulse front is imperfect due to the tilted geometry inherent in this imaging scheme [20]. To mitigate these problems, the use of a stair-step mirror has been proposed [20]. This generates beamlets whose spot size $w_0$ in the image plane is large compared to the pump pulse wavelength λ. Moreover, the Rayleigh range $z_R = \pi w_0^2 \lambda^{-1}$ remains independent of the pulse duration. Thus, beamlets can propagate over a long distance $2z_R$ without appreciable divergence.

To demonstrate this point, we capture the image of the pump beam at different positions along the pump propagation direction with a CCD camera (where the position at 0 mm is the position at the center of the LN crystal). In Fig.3 (a), (b) and (c), we can clearly observe the image of the stepped mirror, which is sharp over a distance of 2 mm. At the 0 mm position, the dimension of the beam is 0.6 mm FWHM along the x-direction and 2.6 mm FWHM along the y-direction. At the maximum pumping energy (208 µJ), the fluence is 17 mJ/cm$^2$, which is more than 40 times below the damage threshold of LN [24]. Furthermore, from the pump beam profiles along the z-direction, we deduced a beam divergence of ~1°. Thus, THz generation is possible over a large portion of the LN due to the large effective interaction length with the slightly varying wavefront tilt angle allowing for efficient and optimal THz generation [20,29]. Furthermore, to validate that the OR process is efficient, we measured the spectrum of the laser beam before and after the LN crystal (see Fig. 3 (e)) and thus observed a significant redshift of the spectrum for the maximum energy of the pump optical pulse (at 208 µJ). This shift enables a cascade effect of optical and THz photons [35]. Thus, one pump photon participates in the generation of several THz photons, which should broaden the generated THz bandwidth.

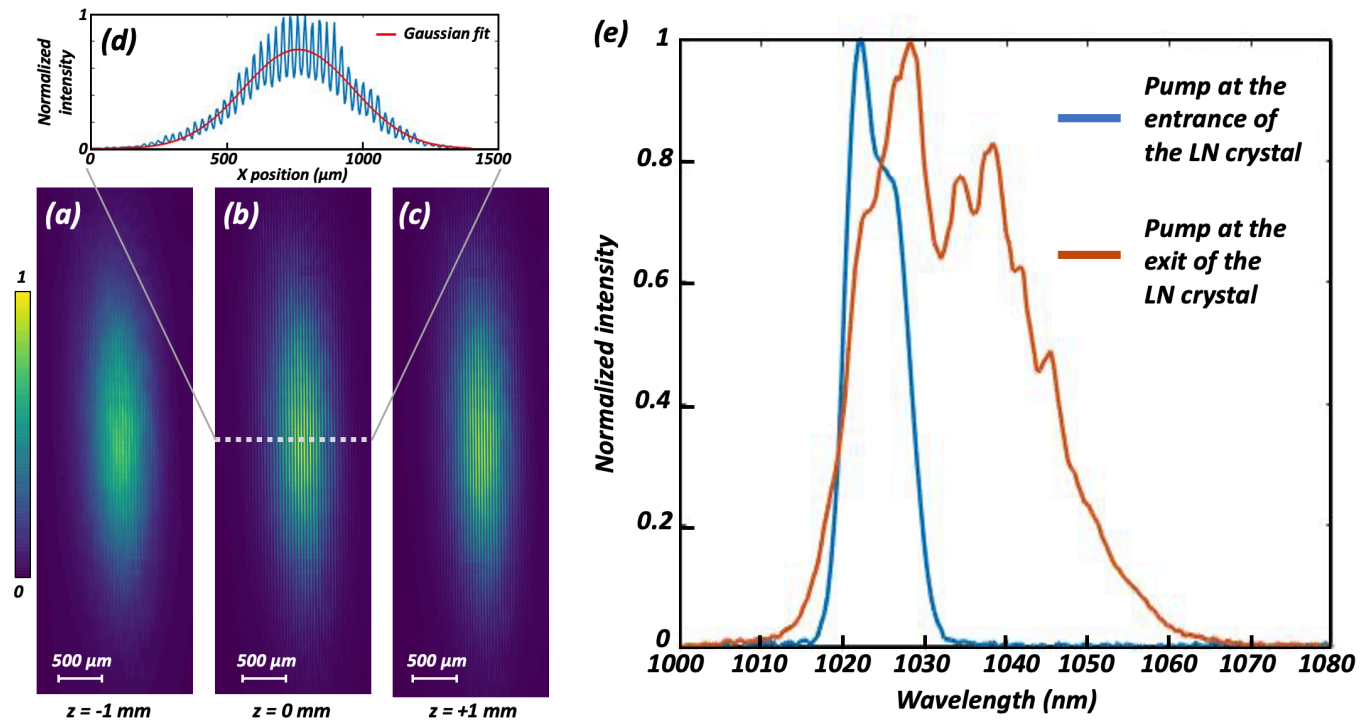

**Fig. 3.** (a) Pump spot image (a) 1 mm before the focus position, (b) at the focus position and (c) 1 mm after the focus position. (d) horizontal profile of the pump spot at the focus position. (e) Normalized pump spectra at the entrance and the exit of the LN crystal after OR process.

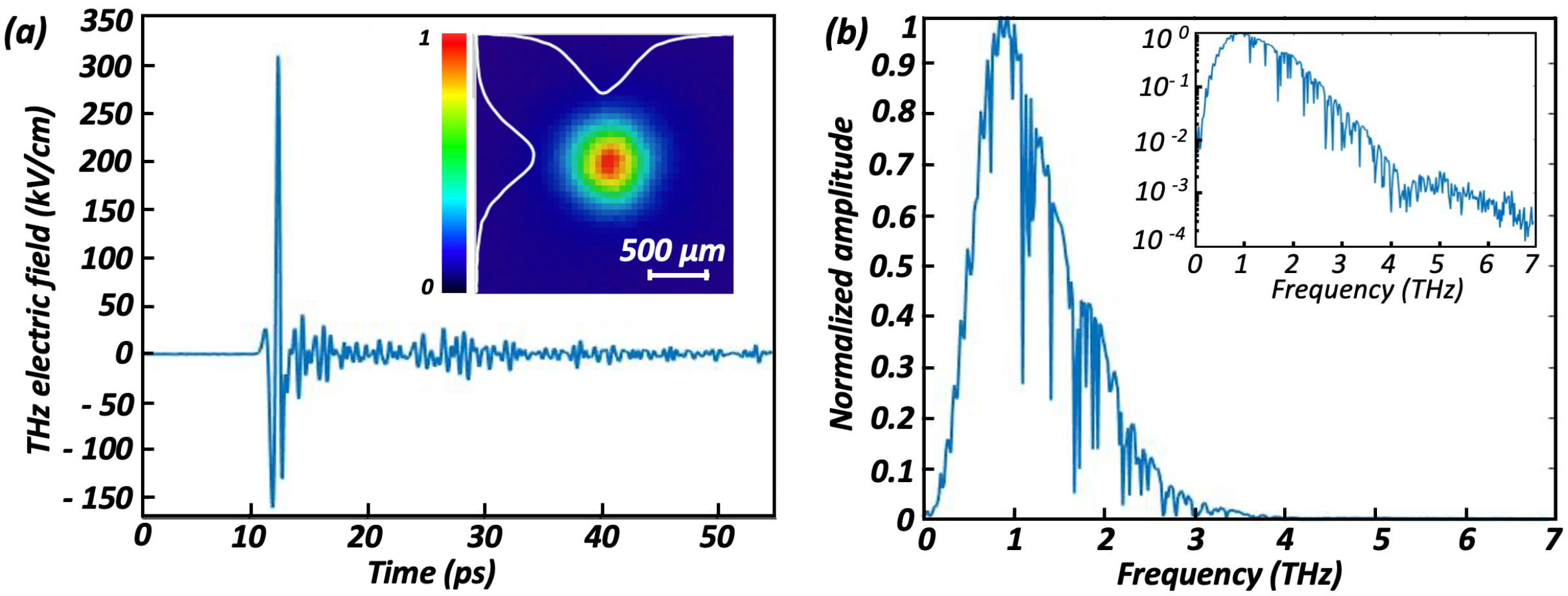

**Fig. 4.** (a) Temporal evolution of the generated THz pulse detected by EO sampling, with the image of the THz spot measured at the focus of the off-axis parabolic mirror $OAP_3$ in the inset. (b) Normalized spectrum of the generated THz pulse, with the normalized spectrum in logarithm scale in the inset.

## 3. RESULTS

The generated THz power was measured using a calibrated pyroelectric detector from GentecEO (THZ5I-BL-BNC) at the focal point of the third off-axis mirror ($OAPM_3$ in Fig. 1). We added four high-resistivity silicon wafers and two undoped germanium wafers to avoid detector saturation. According to the transmission factors of the 6 filters and taking into account the echos from the filters, we measured a maximum THz power of 53 mW for an equivalent energy of 2.1 µJ per pulse. This measurement was obtained for a pumping energy of 208 µJ. This THz output energy corresponds to an optical-THz efficiency of 1%. Finally, taking into account the THz spot size, the pulse duration (estimated at 0.9 ps) and the THz energy, we calculate a peak intensity equal to ~1 GW/cm$^2$.

Figure (4) shows (a) the measured time trace of the THz waveform in an unpurged environment, and (b) the associated spectrum in amplitude in a linear scale, with a logarithmic scale in the inset. The THz waveform is single-cycle and we can clearly distinguish the ringing due to water absorption. We note that the scan range is 43 ps after the main pulse and still no echo of the THz pulse is observed, owing to the 1 mm LN substrate. In the frequency domain, the spectrum ranges from 0.1 to 4 THz and its peak amplitude is located at 0.9 THz. Interestingly, the frequency bandwidth at FWHM covers a range of 0.55-1.6THz with a spectral resolution of 18 GHz. This is, to our knowledge, the largest THz bandwidth ever reported for a source of intense THz pulses from TPFP in an LN crystal at room temperature. This demonstration is in good agreement with a scaling of the result presented in reference [33] and confirms the importance of the cascade effect during the OR process [35].

The inset in Fig. 4 (a) also shows an image of the THz spot size taken at the position of the focus of the third OAPM, with a pyroelectric camera (Electro-physical model PV-320). The dimensions of the THz spot at FWHM are 540 µm in the x-direction and 560 µm in the y-direction. The profile is Gaussian and the spot size shows no sign of ellipticity. It should be noted that the size of the THz spot is not limited by diffraction, which allows to anticipate better performances by slightly changing our THz beam collection system. To fully characterize the source, we measured the THz electric field using a 19 µm thick LN detector and a 21% modulation on the photodiodes was obtained. Using the EO phase delay formula for LN [36], but neglecting the natural birefringence of LN, a peak electric field of about 310 kV/cm was calculated. The THz peak electric field can also be estimated from its energy, duration, and spot size using the formula (2) in Ref. [37]. With this method, we estimate a value of 860 kV/cm, considering a pulse duration of 0.9 ps and a THz spot size of 240 µm$^2$. This method is widely used for the characterization of intense THz sources [22,23,38]. Nevertheless, it easily overestimates the real performance. This can be due to the multiple error

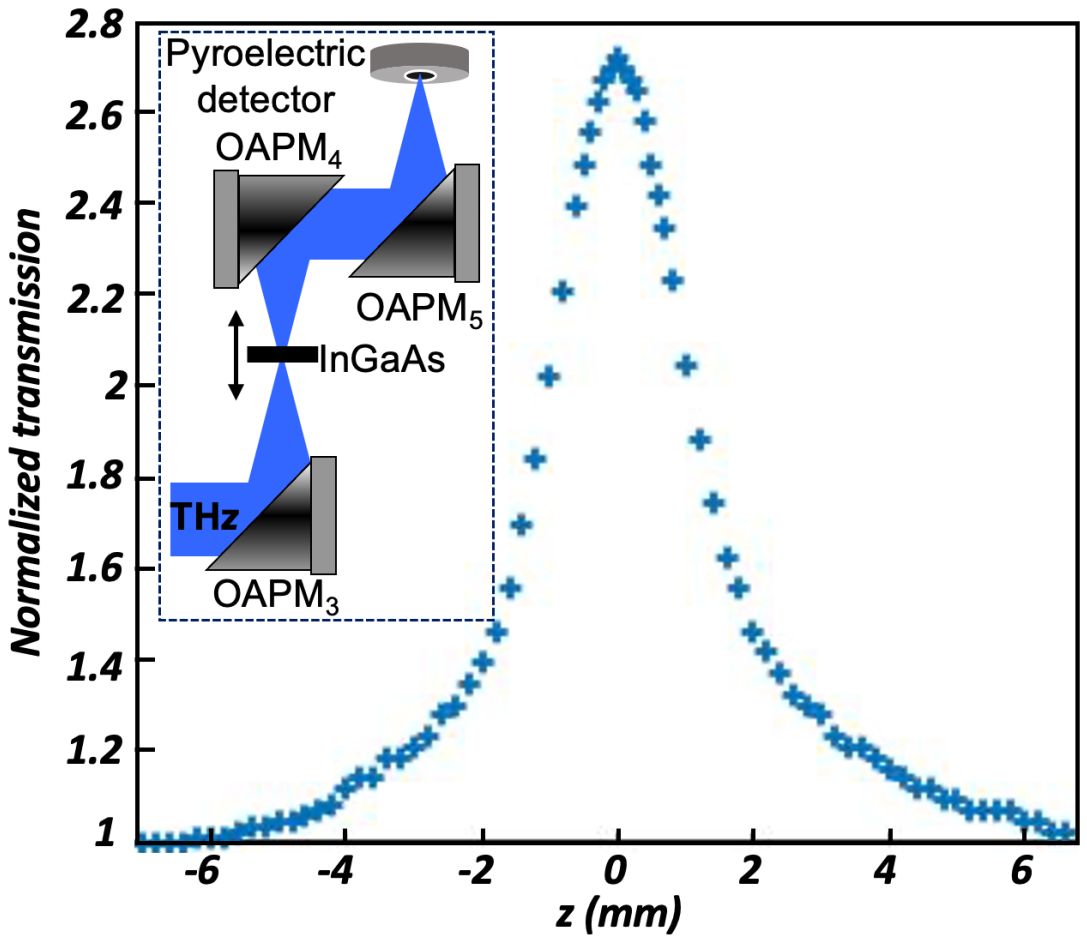

***Fig. 5.*** *Normalized THz transmission through the InGaAs sample as a function of z position [40]. Inset: Experimental setup for the Z-scan measurement with InGaAs sample: $OAPM_3$ of 2" reflected focal length; $OAPM_4$ of 2" reflected focal length; $OAPM_5$ = of 4" reflected focal length.*

**Table 1.** Summary of the performances of some of the recent LN sources based on a TPFP configuration.

| | Input pump laser parameters | | | | | Output characteristics of the generated THz pulses | | | |
|---|---|---|---|---|---|---|---|---|---|
| Ref. | $\lambda_0$ (nm) | $\tau_0$(fs) | PRF (kHz) | $W^L$(mJ) | $P^L(W)$ | $W^{THz}$ (μJ) | $P^{THz}$(mW) | $\eta^{THz}$(%) | $E^{THz}$ (kV/cm) |
| Hirori *et al.* (2011) [14] | 780 | 85 | 1 | 4 | 4 | 3 | 3 | 0.1 | 1200 |
| Fülop *et al.* (2014)[18] | 1030 | 785 | 0.01 | 60 | 0.6 | 463 | 4 | 0.77 | 650 |
| Ofori-Okai *et al.* (2016) [20] | 800 | 70 | 1 | 1.5 | 0.95 | 2.1 | 2.1 | 0.21 | 375 |
| Meyer *et al.* (2020) [25] | 1030 | 550 | 13300 | 0.009 | 123 | 0.005 | 66 | 0.056 | 16.7 |
| Kramer *et al.* (2020) [26] | 1030 | 70 | 100 | 7 | 700 | 1.44 | 144 | 0.042 | 150 |
| Zhang *et al.* (2021) [22]* | 800 | 30 | 0.01 | 500 | 5 | 1400 | 14 | 0.7 | 6300 |
| **This work** | **1024** | **280** | **25** | **0.4** | **10** | **2.1** | **53** | **1** | **310** |

** = cryogenically cooling of the LN; $\lambda_0$= laser central wavelength; $\tau_0$= laser pulse duration; PFR = laser pulse frequency rate; $W^L$ = laser energy; $P^L$ = average laser power;* $W^{THz}$ = THz energy; $P^{THz}$ = average THz power; $\eta^{THz}$ = optical-to-THz conversion efficiency; $E^{THz}$ = electric field strength at the THz peak position.

factors in estimating energy, pulse duration and real area of the THz image. Therefore, we have only considered the lowest evaluation obtained with the method based on the EO modulation of the signal, that is 310 kV/cm at the peak.

Finally, to be sure about the magnitude of the peak electric field of our source, we performed a nonlinear THz experiment with a 500 nm thick n-doped InGaAs crystal (with a carrier density of $2\times10^{18}$ $cm^{-1}$) deposited on a semi-insulating InP substrate [39-41]. The open aperture Z-scan experiment consists of measuring the THz transmission through the InGaAs as a function of its relative position with respect to the THz focal point. The transmitted THz power is measured with the pyroelectric detector. Figure 5 shows the normalized nonlinear transmission as a function of the sample position relative to the focal point with the inset showing the experimental setup. A maximum transmission enhancement factor of 2.7 is obtained at the focus. This value exceeds that reported in the past with the use of a 230 kV/cm THz pulse [41], indicating that our THz electric field evaluation is reasonable.

## 4. CONCLUSION

In summary, we have demonstrated a new intense and powerful THz source that is highly scalable using a 10 W industrial fs laser. Along with this demonstration, we introduced a new pulse compression technique that is compatible with the use of a long Fourier-limited pump pulse duration for efficient THz wave generation. Using less than 1 μJ of probe pulse energy in a simple pulse compression scheme, the THz electric field can be measured with a 75 fs probe beam.

In order to appreciate these combined achievements, we summarize in Table 1 some of the most recent intense THz sources reported to date using the TPFP method in an LN crystal, along with their main characteristics. Compared to reference [26], our source requires 70 times less input laser power to obtain just under twice their average THz power. More importantly, with 17 times less input energy per pump pulse, we obtained twice as much THz intensity, thanks to our high THz conversion $\eta^{THz}$ efficiency of 1%.

Finally, with the use of this highly efficient and accessible system, our demonstration clearly opens the door to high repetition rate nonlinear THz science as well as a wide range of applications outside the field of ultrafast nonlinear spectroscopy (e.g., requiring high brightness at a high repetition rate), which was previously difficult to access for the vast majority of scientists.

**Funding.** F.B. gratefully acknowledges financial support from NSERC (2016-05020) and Canada Research Chair in THz technology.

**Disclosures.** “The authors declare no conflicts of interests.”

**Data availability.** Data underlying the results presented in this paper are not publicly available at this time but may be obtained from the authors upon reasonable request.

## References

1. H. Y. Hwang, S. Fleischer, N. C. Brandt, B. G. Perkins, M. Liu, K. Fan, A. Sternbach, X. Zhang, R. D. Averitt, and K. A. Nelson, "A review of non-linear terahertz spectroscopy with ultrashort tabletop-laser pulses," J. Mod. Opt. 62, 1447–1479 (2015).
2. H. A. Hafez, X. Chai, A. Ibrahim, S. Mondal, D. Férachou, X. Ropagnol and T. Ozaki, "Intense terahertz radiation and their applications," Journal of Optics 18, 093004 (2016).
3. D. Zhang, A. Fallahi, M. Hemmer, X. Wu, M. Fakhari, Y. Hua, H. Cankaya, A.-L. Calendron, L. E. Zapata, N. H. Matlis, and F. X. Kärtner, "Segmented terahertz electron accelerator and manipulator (STEAM)," Nat. Photonics 12, 336–342 (2018).

4. D. Matte, N. Chamanara, L. Gingras, L. P. R. de Cotret, T. L. Britt, B. J. Siwick, and D. G. Cooke, "Extreme lightwave electron field emission from a nanotip," Phys. Rev. Res. 3, 013137 (2021).
5. T. Arikawa, T. Hiraoka, S. Morimoto, F. Blanchard, S. Tani, T. Tanaka, K. Sakai, H. Kitajima, K. Sasaki, and K. Tanaka, "Transer of orbital angular momentum of light to plasmonic excitations in metamaterials," Science Advances 6, eaay1977 (2020).
6. M. Liu, H. Y. Hwang, H. Tao, A. C. Strikwerda, K. Fan, G. R. Keiser, A. J. Sternbach, K. G. West, S. Kittiwatanakul, J. Lu, S. A. Wolf, F. G. Omenetto, X. Zhang, K. A. Nelson, and R. D. Averitt, "Terahertz-field-induced insulator-to-metal transition in vanadium dioxide metamaterial," Nature 487, 345–348 (2012).
7. X. Li, T. Qiu, J. Zhang, E. Baldini, J. Lu, A. M. Rappe, and K. A. Nelson, "Terahertz field-induced ferroelectricity in quantum paraelectric SrTiO3," Science 14, 1079-1082 (2019).
8. H. A. Hafez, S. Kovalev, J.-C. Deinert, Z. Mics, B. Green, N. Awari, M. Chen, S. Germanskiy, U. Lehnert, J. Teichert, Z. Wang, K.-J. Tielrooij, Z. Liu, Z. Chen, A. Narita, K. Müllen, M. Bonn, M. Gensch, and D. Turchinovich, "Extremely efficient terahertz high-harmonic generation in graphene by hot Dirac fermions," Nature 561, 507–511 (2018).
9. J. A. Fülöp, S. Tzortzakis, and T. Kampfrath, "Laser-Driven Strong-Field Terahertz Sources," Adv. Opt. Mater. 8, 1900681 (2020).
10. J. Hebling, G. Almasi, I. Z. Kozma, and J. Kuhl, "Velocity matching by pulse front tilting for large-area THz-pulse generation," Opt. Express 10, 1161 (2002).
11. J. Hebling, K.-L. Yeh, M. C. Hoffmann, B. Bartal, and K. A. Nelson, "Generation of high-power terahertz pulses by tilted-pulse-front excitation and their application possibilities," J. Opt. Soc. Am. B 25, B6 (2008).
12. K.-L. Yeh, M. C. Hoffmann, J. Hebling, and K. A. Nelson, "Generation of 10 μJ ultrashort terahertz pulses by optical rectification," App. Phys. Lett. 90, 171121 (2007).
13. M. C. Hoffmann, K.-L. Yeh, J. Hebling, and K. A. Nelson, "Efficient terahertz generation by optical rectification at 1035 nm," Opt. Express 15, 11706 (2007).
14. H. Hirori, A. Doi, F. Blanchard, and K. Tanaka, "Single-cycle terahertz pulses with amplitudes exceeding 1 MV/cm generated by optical rectification in LiNbO3," Appl. Phys. Lett. 98, 091106 (2011).
15. J. A. Fülöp, L. Pálfalvi, M. C. Hoffmann, and J. Hebling, "Towards generation of mJ-level ultrashort THz pulses by optical rectification," Opt. Express 19, 15090-15097 (2011).
16. J. A. Fülop, L. Palfalvi, S. Klingebiel, G. Almasi, F. Krausz, S. Karsch, and J. Hebling, "Generation of sub-mJ terahertz pulses by optical rectification," Opt. Lett. 37, 557-559 (2012).
17. F. Blanchard, X. Ropagnol, H. Hafez, H. Razavipour, M. Bolduc, R. Morandotti, T. Ozaki, and D. G. Cooke, "Effect of extreme pump pulse reshaping on intense terahertz emission in lithium niobate at multimilliJoule pump energies," Opt. Lett. 39, 4333 (2014).
18. J.-A. Fülop, Z. Olimann, Cs. Lombosi, C. Skrobol, S. Klingebiel, L. Palfalvi, F. Krausz, S. Karsch, and J. Hebling, "Efficient generation of THz pulses with 0.4 mJ energy," Opt. Express 22, 20155-20163 (2014).
19. W. R. Huang, S.-W. Huang, E. Granados, K. Ravi, K.-H. Hong, L. E. Zapata, and F. X. Kärtner, "Highly efficient terahertz pulse generation by optical rectification in stoichiometric and cryo-cooled congruent lithium niobate," J. Mod. Opt. 62, 1486–1493 (2015).
20. B. K. Ofori-Okai, P. Sivarajah, W. Ronny Huang, and K. A. Nelson, "THz generation using a reflective stair-step echelon," Opt. Express 24, 5057 (2016).
21. X.-J. Wu, J.-L. Ma, B.-L. Zhang, S.-S. Chai, Z.-J. Fang, C.-Y. Xia, D.-Y. Kong, J.-G. Wang, H. Liu, C.-Q. Zhu, X. Wang, C.-J. Ruan, Y.-T. Li, ''Highly efficient generation of 0.2 mJ terahertz pulses in lithium niobate at room temperature with sub-50 fs chirped Ti:sapphire laser pulses'' Opt. Express 26, 7107-7116 (2018).
22. B. Zhang, Z. Ma, J. Ma, X. Wu, C. Ouyang, D. Kong, T. Hong, X. Wang, P. Yang, L. Chen, Y. Li, and J. Zhang, "1.4-mJ high energy terahertz radiation from lithium niobates," Laser Photonics Rev. 15, 2000295, (2021).
23. Q. Tian, H. Xu, Y. Wang, Y. Liang, Y. Tan, X. Ning, L. Yan, Y. Du, R. Li, J. Hua, W. Huang, and C. Tang, "Efficient generation of a high-field terahertz pulse train in bulk lithium niobate crustals by optical rectification," Opt. Express 29, 9624-9634 (2021).
24. F. Bach, M. Mero, M.-H. Chou, and V. Petrov, "Laser induced damage studies of LiNbO_3 using 1030-nm, ultrashort pulses at 10-1000 kHz," Opt. Mater. Express 7, 240 (2017).
25. F. Meyer, T. Vogel, S. Ahmed, and C. J. Saraceno, "Single-cycle, MHz repetition rate THz source with 66 mW of average power," Opt. Lett. 45, 2494 (2020).
26. P. L. Kramer, M. K. R. Windeler, K. Mecseki, E. G. Champenois, M. C. Hoffmann, and F. Tavella, "Enabling high repetition rate nonlinear THz science with a kilowatt-class sub-100 fs laser source," Opt. Express 28, 16951 (2020).
27. M. Abdo, S. Sheng, S. Rolf-Pissarczyk, L. Arnhold, J. A. J. Burgess, M. Isobe, L. Malavolti, and S. Loth, "Variable repetition rate THz source for ultrafast scanning tunneling microscopy," ACS Photonics 8, 702-708 (2021).
28. J. Henrich, S. Butcher, and M. Arrigoni, "Ultrafast lasers: Trends in femtosecond amplifiers - Ti:sapphire vs. ytterbium," LaserFocusWorld, Feb 18th (2020).
29. L. Wang, G. Toth, J. Hebling, and F. Kärtner, "Tilted-pulse-front schemes for terahertz generation," Laser & Photonics Rev. 14, 2000021 (2020).
30. K. Murate, M. J. Roshtkhari, X. Ropagnol, and F. Blanchard, "Adaptive spatiotemporal optical pulse front tilt using a digital micromirror device and its terahertz application," Opt. Lett. 43, 2090-2093 (2018).
31. Q. Wu and X.-C. Zhang, "Ultrafast electro-optic field sensors," Appl. Phys. Lett. 68, 1604 (1996).
32. F. Blanchard, A. Doi, T. Tanaka, H. Hirori, H. Tanaka, Y. Kadoya, and K. Tanaka, "Real-time terahertz near-field microscope," Opt. Express 19, 8277–8284 (2011).
33. M. Nagai, E. Matsubara, M. Ashida, J. Takayanagi, and H. Ohtake, "Generation and Detection of THz Pulses With a Bandwidth Extending Beyond 4 THz Using a Subpicosecond Yb-Doped Fiber Laser System," IEEE Trans. THz Sci. Technol. 4, 440–446 (2014).
34. O. E. Martinez, "Pulse distortions in tilted pulse schemes for ultrashort pulses," Optics Communications 59, 229–232 (1986).
35. M. Jewariya, M. Nagai, and K. Tanaka, "Enhancement of terahertz wave generation by cascaded $X^{(2)}$ processes in LiNbO3," JOSA B 26, A101-A106 (2009).
36. F. Amirkhan, R. Sakata, K. Takiguchi, T. Arikawa, T. Ozaki, K. Tanaka, and F. Blanchard, "Characterization of thin-film optical properties by THz near-field imaging method," JOSA B 36, 2593-2601 (2019).
37. M. Reid, and R. Fedosejevs, "Quantitative comparison of THz emission from (100) InAs surfaces and GaAs large-aperture photoconductive switch at high fluences," Appl. Opt.44, 149 (2004).
38. F. Blanchard, L. Razzari, H. C. Bandulet, G. Sharma, R. Morandotti, J. C. Kieffer, T. Ozaki, M. Reid, H. F. Tiedje, H. K. Haugen, and F. A. Hegmann, "Generation of 1.5 μJ single-cycle terahertz pulses by optical rectification from a large aperture ZnTe crystal," Opt. Express 15, 13212 (2007).
39. L. Razzari, F. H. Su, G. Sharma, F. Blanchard, A. Ayesheshim, H.-C. Bandulet, R. Morandotti, J.-C. Kieffer, T. Ozaki, M. Reid, and F. A. Hegmann, "Nonlinear ultrafast modulation of the optical absorption of intense few-cycle terahertz pulses in n -doped semiconductors," Phys. Rev. B **79**, 193204 (2009).
40. F. Blanchard, D. Golde, F. H. Su, L. Razzari, G. Sharma, R. Morandotti, T. Ozaki, M. Reid, M. Kira, S. W. Koch, and F. A. Hegmann, "Effective mass anisotropy of hot electrons in nonparabolic conduction bands of n-doped InGaAs film using ultrafast terahertz pump-probe techniques," Phys. Rev. Lett. 107, 107401 (2011).
41. A. Rovere, Y.-G. Jeong, R. Piccoli, S.-H. Lee, S.-C. Lee, O.-P. Kwon, M. Jazbinsek, R. Morandotti, and L. Razzari, "Generation of high-field terahertz pulses in an HMQ-TMS organic crystal pumped by an ytterbium laser at 1030 nm," Opt. Express **26**, 2509 (2018).